\documentclass[aps,prl,twocolumn,superscriptaddress,nofootinbib,longbibliography]{revtex4-2}

\pdfoutput=1
\usepackage{mathtools,stmaryrd}

\usepackage{graphicx}
\usepackage{amsmath,amssymb,amsfonts}
\usepackage[colorlinks,citecolor=blue,linkcolor=blue,urlcolor=blue, breaklinks=true]{hyperref}
\usepackage{color}
\usepackage[normalem]{ulem}

\usepackage[T1]{fontenc}

\usepackage{physics} 
\usepackage[percent]{overpic} 
\usepackage{bm}

\usepackage[english]{babel}

\newtheorem{theorem}{Theorem}
\newtheorem{lemma}[theorem]{Lemma}

\newtheorem{problem}[theorem]{Problem}

\usepackage{xcolor}
\usepackage{xspace}
\usepackage{ifthen} 

\newcommand{\showcomments}{true}

\newcommand{\marios}[1]%
{\ifthenelse{\equal{\showcomments}{true}}%
{{\color{orange}{\small \textbf{M:} #1}}}{\xspace}}%

\newcommand{\aristos}[1]%
{\ifthenelse{\equal{\showcomments}{true}}%
{{\color{blue}{\small \textbf{A:} #1}}}{\xspace}}%

\usepackage[normalem]{ulem}

\begin{document}

\title{Incompleteness Theorems for Observables in  General Relativity}

\author{Aristotelis Panagiotopoulos}
\affiliation{Department of Mathematical Sciences, Carnegie Mellon University (CMU), Wean Hall, 5000 Forbes Ave, Pittsburgh, PA 15213}

\author{George Sparling}
\affiliation{Laboratory of Axiomatics, University of Pittsburgh, 301 Thackeray Hall, Pittsburgh, PA 15260}

\author{Marios Christodoulou}
\affiliation{Institute for Quantum Optics and Quantum Information (IQOQI) Vienna, Austrian Academy of Sciences, Boltzmanngasse 3, A-1090 Vienna, Austria}

\date{\small\today}

\begin{abstract} \noindent 
The quest for complete observables in general relativity has been a longstanding open problem. We employ methods from  descriptive set theory to show that no complete observable 
on rich enough collections of spacetimes 
is Borel definable. In fact, we show that it is consistent with the Zermelo-Fraenkel and Dependent Choice axioms that  no  complete observable for rich  collections of spacetimes exists whatsoever. In a nutshell, this implies that the Problem of Observables is to `analysis' what the Delian Problem  was to `straightedge and compass'. Our results remain true even after restricting the space of solutions to vacuum solutions. In other words, the issue can be traced to the presence of local degrees of freedom.  We discuss the next steps in a  research program that aims to further uncover this novel connection between theoretical physics and descriptive set theory.
\end{abstract}

\maketitle

From Einstein's century old hole--argument--paradox \cite{Ein1,Ein2,Norton}, to the contemporary programs for quantizing gravity \cite{Old3,Thiemann,Car2001,ashtekar_reuter_rovelli_2015},
the problem of deciding which `functions' of the metric components do not depend on the choice of coordinates  has raised  technical and epistemological difficulties in the theory of  general relativity (GR). This issue  has become known as \emph{the problem of observables}. 

The quest for complete observables --- observables which can discern between any pair of diffeomorphically inequivalent spacetimes --- begins in the 1950s \cite{Old0,Old1,Old2,BG,Obse,Old3}, and is often discussed in the context of more modern approaches \cite{New0,New1,New2,New3,New4,New5}.  While one can  tailor observables for  special families of spacetimes  \cite{ADM,Dust1,Dust2},   no non--trivial (non--constant) observable supported on the collection of all spacetimes has been reported. This, despite a seven--decades--long search since Bergmann famously stated the issue \cite{Old0,Old1,Old2,BG}. 
     The question arises: why this state of affairs? Notwithstanding some interesting  partial negative results in the Hamiltonian formulation of the problem \cite{Torre, Neg}, a conclusive  result 
   which identifies the root of the issue has remained elusive. 

In this letter, we employ methods from  descriptive set theory to prove a rather conclusive negative result for the definability of complete observables, at least when no significant constraints on the space of solutions are imposed: \emph{there is  no constructive way to build complete observables for full general relativity}. We trace the root cause of this incompleteness phenomenon to a certain  
ergodic-theoretic behaviour that general covariance exhibits on any `rich enough' collection of spacetimes.

Our results do not imply that a theory of quantum gravity cannot be  based on  definable observables of some kind. Rather, they highlight some of the difficulties when considering large and diverse collections $\mathcal{S}$ of spacetimes, all at once --- a problem not unique to GR, but potentially to any other physical theory with a large  symmetry group. 
Indeed, a takeaway is that a physical theory can be extremely useful even when the full space of solutions is too large to admit definable complete observables.

Theorems \ref{T:1} and \ref{T:3} provide the precise statements. Both theorems hold for any collection of spacetimes $\mathcal{S}$  that is \emph{rich} ---a  technical term  that we define below. In particular, they both  hold when  $\mathcal{S}$ is the collection of all  Lorentzian manifolds of dimension $d$ for any fixed $d\geq 2$.

\begin{theorem}\label{T:1}
No concrete observable $f\colon \mathcal{S}\to R$ is both
complete and  Borel definable.
\end{theorem}

 The terms appearing in the statement of Theorem \ref{T:1} will be  defined below. In plain language:  completeness requires that $f$   distinguishes any two diffeomorphically inequivalent spacetimes by assigning to  them different values; concreteness requires that $f$ takes concrete  objects as values, e.g.\;real numbers, invariant scalars, etc; Borel definability requires that $f$ is given by some formula expressible in the language of analysis.

Theorem \ref{T:1} shows that it is as futile to seek an analytic description for a complete observable, as  trying to construct  $\sqrt[3]{2}$  using straightedge and compass. This is not to say that complete observables do not `exist'. In the extremely abstract sense allowed when utilising the Axiom of Choice (AC), complete observables do exist. However, for a mathematical object to be useful in doing  physics, it should also be amenable to some kind of description with analytic tools. In a sense, when an object exists only by the power of AC, then for what concerns physics it is as useful as if it did not exist. From this point of view, the following is even more troubling.

\begin{theorem}\label{T:3} 
The statement ``no complete concrete observable for $\mathcal{S}$ exists'' is consistent with $\mathrm{ZF}+\mathrm{DC}$. 
\end{theorem}

Here  ZF stands for the usual  Zermelo-Fraenkel axioms of set theory and DC stands for the axiom of Dependent Choice: a  `fragment' of AC  that is needed even for basic real analysis on the Euclidean space. Theorem \ref{T:3} is  proved in ZF+AC (ZFC)  and it highlights the non-constructive nature of complete observables: any mathematical proof of the statement that complete observables merely `exist', has to make use of the  `full' strength of AC.

 Importantly, \emph{both  theorems above hold even if we restrict $\mathcal{S}$ to be  the family of  vacuum solutions\footnote{ Solutions with vanishing stress--energy tensor and cosmological constant $\Lambda=0$. For $\Lambda\neq 0$ we refer to the discussion.} on $\mathbb{R}^4$.}
 That is, the problem can be traced to the  local degrees of freedom present in the vacuum theory. This is a feature that is particular to 4 dimensions, and perhaps higher.  Indeed, it is in sharp contrast to the vacuum theory on $\mathbb{R}^3$,  which trivially admits complete observables,  as its only geodesically complete solution is the Minkowski spacetime. Another important point is that the above results
 immediately extend to incompleteness theorems  for \emph{countable families} of concrete and definable observables. 

 \medskip

Theorems \ref{T:1} and \ref{T:3} follow from  Lemma \ref{Lemma:1}, a stronger but more technical version of Theorem \ref{T:1}. All three results are proved for an arbitrary space of solutions $\mathcal{S}$ which is `rich', see  below. In Theorem \ref{T:Examples}
we show that the family of gravitational plane waves is rich. This implies that the vacuum sector of solutions is also rich. The proof of Theorem \ref{T:Examples} is given in the Supplementary Material, where we also show that the family of Robertson-Walker spacetimes in any dimension   $d\geq 2$ is  rich.

Our results do not imply that all questions regarding the definability of observables have been addressed. In closing, we discuss their reach and speculate on strategies for trying to circumvent incompleteness, by relaxing the notion of `observables' or by 
restricting the space $\mathcal{S}$ of `acceptable' solutions. We put forth a series of open problems which aim to form  the  backbone of a research program to  determine the  intrinsic complexity of general covariance and help identify quantization procedures that  could be  implemented constructively.

\medskip
\emph{The problem of observables---} Originating in the work of Bergmann   \cite{Old0,Old2,BG}, the problem of observables refers to the problem of identifying those  ``\emph{functions (or functionals) of field variables that are invariant with respect to coordinate transformations}'' \cite{Obse}.
Formally, an observable for a collection $\mathcal{S}$ of metric component fields is any function  $f\colon \mathcal{S}\to R$  to a set $R$, so that for all  $g_{\mu\nu}, \widetilde{g}_{\rho\sigma}\in\mathcal{S}$
\begin{equation}\label{EQ:Invariant}
 g_{\mu\nu}\simeq_{\mathrm{diff}}\widetilde{g}_{\rho\sigma} \implies f(g_{\mu\nu})=f(\widetilde{g}_{\rho\sigma}).
\end{equation}
We write 
$\widetilde{g}_{\rho\sigma}\simeq_\mathrm{diff} g_{\mu\nu}$  whenever there exists a smooth  change  of coordinates $\widetilde{x}^{\xi}=\widetilde{x}^{\xi}(x^{\eta})$ so that
\begin{equation}\label{EQ:GC}
g_{\mu\nu}(x^{\eta})=\frac{\partial \widetilde{x}^{\rho} }{\partial x^{\mu}} \frac{\partial \widetilde{x}^{\sigma} }{\partial x^{\nu}} \widetilde{g}_{\rho\sigma}(\widetilde{x}^{\xi}).
\end{equation}

The  goal in Bergmann's program was to piece together a complete family of observables. That is, enough observables to tell apart  different geometries represented in $\mathcal{S}$, similarly to how  Komar mass  \cite{Komar}  classifies  Schwarzschild spacetimes.  Since the notions of concretness and defibability below are closed under countable products, we can always  replace a  list $f_1,\ldots f_n,\ldots$ of observables with a single observable $f=\otimes_n f_n$.  Hence,
 it suffices to consider completeness in the context of a single observable.

\medskip
\emph{Completeness---}  An observable $f\colon \mathcal{S}\to R$ is complete for $\mathcal{S}$ if, for all $g_{\mu\nu}, \widetilde{g}_{\rho\sigma}\in\mathcal{S}$, we can strengthen (\ref{EQ:Invariant}) to
\begin{equation}\label{EQ:Complete}
g_{\mu\nu}\simeq_{\mathrm{diff}}\widetilde{g}_{\rho\sigma} \iff f(g_{\mu\nu})=f(\widetilde{g}_{\rho\sigma}).
\end{equation}
Without imposing any further restrictions on the `concreteness' of the range $R$ and the `definability' of $f$, any  space of solutions $\mathcal{S}$ admits a complete observable.  For example, one can always take $R$ to be the `abstract' collection of all  equivalence classes represented in $\mathcal{S}$
\begin{equation}\label{EQ:Abstract}
[g_{\mu\nu}]_{\mathrm{diff}} :=  \{\widetilde{g}_{\rho\sigma}\in \mathcal{S}\colon \widetilde{g}_{\rho\sigma}\simeq_\mathrm{diff} g_{\mu\nu}\},
\end{equation}
and consider the complete observable that is given by
the assignment  $g_{\mu\nu}\mapsto [g_{\mu\nu}]_{\mathrm{diff}}$. 
Or, one can take  $R=\mathbb{R}$ to be the more `concrete' space of all real numbers, and use AC to build a complete $\mathbb{R}$-valued observable.

To rule out such extreme `solutions' to the problem of observables, we will next require that  observables are concrete and definable. For these notions, as well as for a few more technical points later on, we will need some nomenclature  from descriptive set theory \cite{Kechris}.

\medskip
\emph{Elements of descriptive set theory}--- Let $X$ be a topological space and let  $A\subseteq X$. Then, $A$ is nowhere dense if the complement of its closure is dense in $X$; meager if it is a countable union of nowhere dense sets;  comeager if its complement is meager; Borel if it is in the smallest $\sigma$-algebra of subsets of $X$ that contains the open sets; Baire-measurable if it is in the smallest   $\sigma$-algebra of subsets of $X$ that contains both the open and the nowhere dense subsets of $X$.  A map $f\colon X\to Y$ between topological spaces is Borel --- respectively, Baire-measurable ---   if so is $f^{-1}(U)$, for every open $U\subseteq Y$. We are particularly interested in Polish spaces, where these notions are well behaved. A Polish space is a  topological space $X$ whose topology is separable and completely metrizable.

\medskip
\emph{The  Borel structure on $\mathcal{S}$}---
Let $\mathrm{Ein}(M)$ denote the collection of all smooth spacetimes supported on a smooth manifold $M$. In what follows, we assume that $\mathcal{S}$ is a subset of $\mathrm{Ein}(M)$, for some fixed $M$.
We  denote by $\tau$ the  $C^{\infty}$ compact-open topology on $\mathcal{S}$.  Specifically, let $C^{\infty}(M,N)$ be the Polish space  of all smooth maps $M\to N$ between two manifolds $M,N$ endowed with the $C^{\infty}$ compact-open topology. A basic open  $U_{f,K,n,\varepsilon}\subseteq C^{\infty}(M,N)$ consists of all $g\in C^{\infty}(M,N)$ whose derivatives up to degree $n$ on the compact $K\subseteq M$  are $\varepsilon$-close  with those of $f$ \cite{Hirsch}. 
With the usual identifications we view $\mathcal{S}$ as
 a subset of  $C^{\infty}(M,N)$, where  $N:=(TM\otimes TM)^{*}$. Then $\mathcal{S}$ inherits from $C^{\infty}(M,N)$ the $C^{\infty}$ compact-open topology. This topology induces  on $\mathcal{S}$  the $\sigma$-algebra of  Borel sets on which we base the notion of definable observables  below.

Theorem \ref{T:1} is a statement about the Borel sets on $\mathrm{Ein}(M)$, so it implicitly relies on $\tau$. Note that $\tau$ is a rather weak topology and  of no obvious physical relevance. Here, it is used as a convenient `basis' for spanning the $\sigma$-algebra of Borel sets, which is  a more robust structure.  For example, any stronger Polish topology $\tau'\supseteq \tau$  on $\mathrm{Ein}(M)$ induces the exact same Borel sets and all physical spacetimes, i.e., sets of the form  (\ref{EQ:Abstract}), are themselves Borel\footnote{These follow from \cite[Exercise 15.4]{Kechris} and \cite[Proposition 3.1.10]{Gao}.}. On the other hand, once $\mathcal{S}$ has been established to be rich, the conclusion of Theorem \ref{T:3} is agnostic  both on the topology and the Borel structure on $\mathcal{S}$.

\medskip
\emph{Concreteness---}
An observable $f\colon \mathcal{S}\to R$ is concrete if it takes values in a Polish space. Restricting $R$ to be a Polish space is a generic requirement. For instance, setting $R$ to be either of the Polish spaces $\mathbb{R}$ or $C^{\infty}(M,\mathbb{R})$  we recover  classical definitions of observables \cite{BG,Obse}. However,  our definition of concreteness allows observables to take values in much more general spaces,  as Polish spaces include spaces of distributions, separable Banach spaces,  as well as a vast array of more `exotic' objects like the Cantor set. 
Restricting    $R$ to be a Polish space is  natural as well   from the viewpoint of descriptive set theory, which considers  Polish spaces to be  `well behaved' incarnations of uncountable sets. This is because their points are controlled by a countable dense subset, similarly to how the rationals  control the reals.

\medskip
\emph{Definability---} A concrete observable  $f\colon \mathcal{S}\to R$ is  \emph{Borel definable} if it is a Borel map when $\mathcal{S}$ is endowed with the $C^{\infty}$ compact-open topology. 
These are exactly those observables which admit a description by an explicit  formula in the language of analysis, in the following sense.  

The descriptive  power of analysis is rooted in its ability to implement limiting procedures. 
For example, defining the value $f(g_{\mu\nu})$ of an  ADM observable $f$ requires `taking limits' at  least once, as it is given as the limit of integrals over a sequence of compact regions of the manifold \cite{ADM}.
Maps whose definition relies on limiting procedures  of length two can already be surprisingly complex. For instance,  the characteristic map $\chi_{\mathbb{Q}}\colon \mathbb{R}\to \mathbb{R}$ of the rationals  can be expressed as   $\chi_{\mathbb{Q}}(x)=\lim_{n}\lim_{m} \cos^{2m}(\pi  n!  x)$.

Borel maps are precisely those maps which are attained by allowing
iterations of such limiting procedures for any `number' $\xi$ of times, where $\xi$ ranges over the set  $\omega_1$ of all countable ordinals \cite[Theorem 24.3]{Kechris}.

\medskip
\emph{Rich Families---}  
Theorems \ref{T:1} and \ref{T:3} concern any family of solutions $\mathcal{S}$ which is \emph{rich}. For the definition of this notion we recall a few more  elements from \emph{invariant} descriptive set theory \cite{hjorth, Gao}.

A Polish group $G$ is a topological group  whose topology is Polish. A Polish $G$-space  is   a continuous action $G\curvearrowright X$ of the Polish group $G$ on a  Polish space $X$. The associated \emph{orbit equivalence relation} $\simeq_{G}$ on $X$ is given by setting  $x\simeq_{G}y$  if and only if $x,y$ are in the same orbit, i.e., if $G\cdot x= G\cdot y$. We say that $G\curvearrowright X$ is  \emph{generically ergodic} if: (1) there is $x\in X$, whose orbit  $G\cdot x$ is dense in $X$; (2)  for every $x\in X$, the orbit $G\cdot x$ is meager in $X$.

A family $\mathcal{S}$ of spacetimes is called \emph{rich}, if there exists a generically ergodic Polish $G$-space $G\curvearrowright X$ together with a \emph{Borel  reduction} $r$ from  $(X,\simeq_{G})$ to $(\mathcal{S},\simeq_{\mathrm{diff}})$. That  is, a
Borel map  $r\colon X\to \mathcal{S}$ so that for all
$\alpha,\beta\in X$ we have
\begin{equation}\label{EQ:reduction}
\alpha\simeq_{G}\beta \iff r(\alpha)\simeq_{\mathrm{diff}}r(\beta).
\end{equation}
One way for $\mathcal{S}$ to be rich is if the action $\mathrm{Diff}(M)\curvearrowright \mathcal{S}$
of the diffeomorphism group, implementing \eqref{EQ:GC}, is itself generically ergodic. In this case, it is very difficult to tell different orbits apart as any open set in $\mathcal{S}$  will be intersected by almost every orbit, and the mental picture which  depicts  orbits as `curves' should better be replaced with that of a `knotted ball of yarn'.
That being said, for $\mathcal{S}$ to be rich it is enough for this tangling between orbits to occur just in some `corner' of $\mathcal{S}$.

\medskip
\emph{The family of vacuum solutions---} Before we turn to the proofs of Theorems \ref{T:1} and \ref{T:3}  we would like to establish that  rich families of solutions exist and hence, these theorems are not vacuous. 
To the reader familiar with these arguments, it is probably not that  surprising that rich families exist. Indeed,  without imposing any restrictions on the stress-energy tensors of the members of $\mathcal{S}$, one can simply
concoct rich families of  energy--momentum distributions that  generate  ergodic behaviour within $\mathcal{S}$. An example which illustrates this can be found in the Supplementary Material, where we show that the family of cosmological Robertson-Walker spacetimes \cite{Cosmo} is rich.

Perhaps what is more surprising is that the problem is already present in the vacuum sector. That is, even when all members of $\mathcal{S}$ have a vanishing stress--energy tensor (we assume  the  cosmological constant $\Lambda$ to be zero).

\begin{theorem}\label{T:Examples}
Vacuum solutions on $\mathbb{R}^4$ form a rich family.
\end{theorem}
Theorem \ref{T:Examples} implies that any collection of spacetimes which contains the vacuum solutions on $\mathbb{R}^4$ is rich. In particular, this provides yet another proof that the collection of all spacetimes is rich ---one which does not  rely on the collection of Robertson-Walker spacetimes.

We  now sketch the proof of Theorem \ref{T:Examples}, detailed in the Supplementary Material.
Consider the family $\mathrm{GPW}$ of  gravitational plane waves on $\mathbb{R}^4$. These are all spacetimes which can be written in Brinkmann  form \cite{Brink} as
\begin{equation}\label{EQ:PlaneWaves}
H(u,x,y) du^2 + dudv + dx^2 +dy^2,
\end{equation}
where $H$ is a smooth map that is quadratic in $x,y$ and satisfies  $H_{xx}-H_{yy}=0$. Since  members of $\mathrm{GPW}$ are vacuum solutions  \cite{JEK},  it suffices to see that $\mathrm{GPW}$ is rich.

As a model of generic ergodicity we will use the Bernoulli shift $\mathbb{Z}\curvearrowright X$, where
 $X:=\{0,1\}^{\mathbb{Z}}$ is the space of all integer-indexed  sequences of $0,1$ endowed with the  product topology. The action $\mathbb{Z}\curvearrowright X$ is implemented by $(k,\alpha):=k\cdot \alpha$, where $(k\cdot\alpha)(n)=\alpha(n-k)$. Hence, 
\begin{equation}\label{EQ:ZShift}
\alpha\simeq_{\mathbb{Z}}\beta \iff \exists k\in\mathbb{Z} \; \forall n\in\mathbb{Z} \; \; \alpha(n-k)=\beta(n).
\end{equation}
To see that  $\mathbb{Z}\curvearrowright X$ is generically ergodic, notice that its orbits are countable and
 that for the  random $\alpha\in X$ in the sense of the coin-flip measure,  $\alpha$ admits a dense orbit.

\begin{figure}[h!]

\includegraphics[scale=0.27]{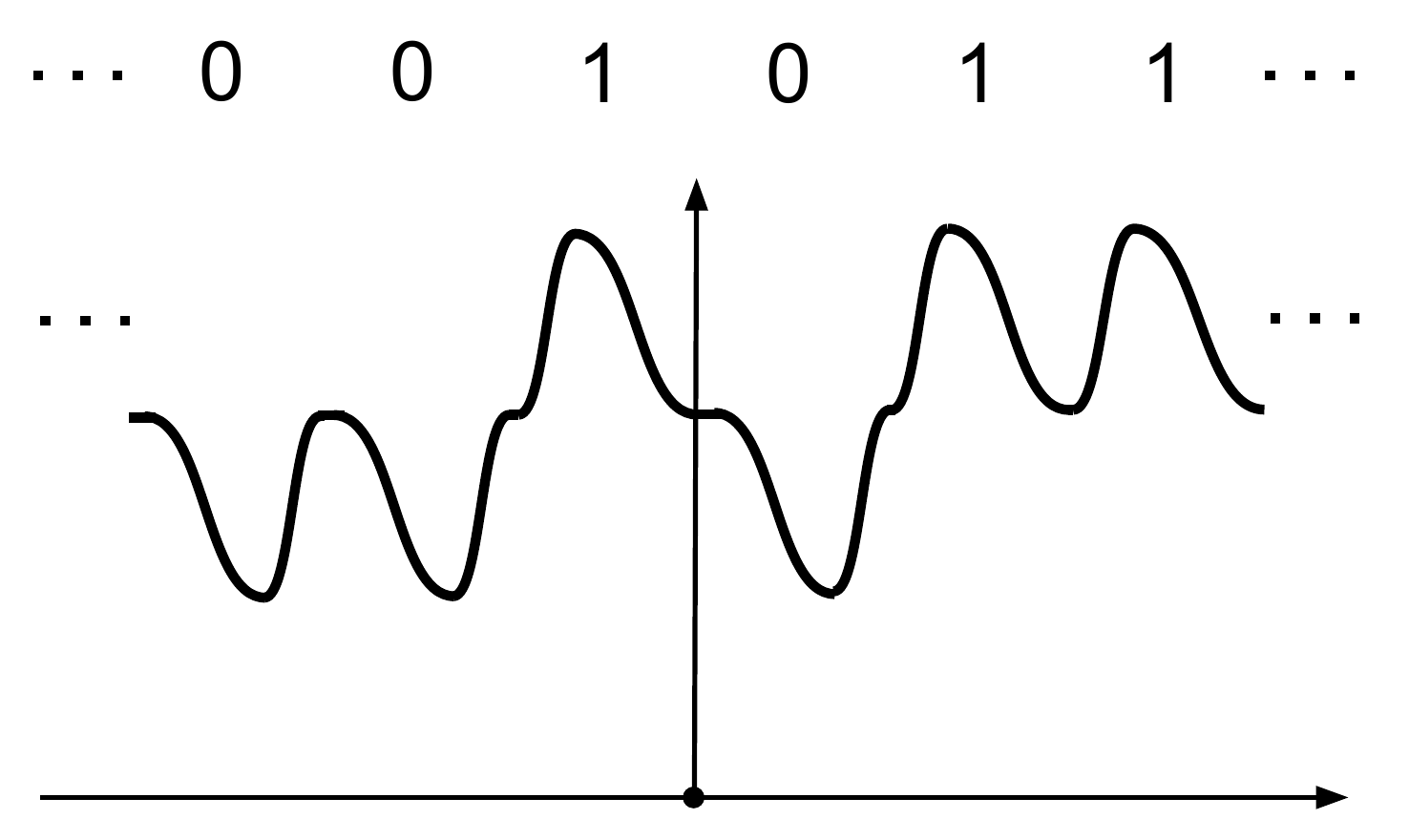}

\caption{To each   $\alpha\in X$ we associate a smooth $W_{\alpha}$}
\label{Fig:1}
\end{figure}

We can now associate a smooth map  $W_{\alpha}\colon \mathbb{R}\to \mathbb{R}$ to each  $\alpha\in X$, so that $W_{\alpha}$ reflects the distribution of $0,1$'s in the sequence $\alpha$; see Figure \ref{Fig:1}. We  define
a Borel reduction $r\colon X\to \mathrm{GPW}$ by setting
 $r(\alpha)$ to be the metric with  $H(u,x,y):=W_{\alpha}(u)xy$ in  (\ref{EQ:PlaneWaves}). The map $r$ is in fact continuous since compact regions of $r(\alpha)$ are determined by finite regions of $\alpha$. It is straightforward  to check that  $r$ satisfies the $(\Longrightarrow)$ direction of (\ref{EQ:reduction}). The $(\Longleftarrow)$ direction of (\ref{EQ:reduction}) also holds and is given in the Supplementary Material. This part  is more technical and it relies on the theory of Lie symmetries of planes waves from  \cite{JEK,Sippel,SKMHH}.

In summary, we showed that the highly tangled orbit structure of the Bernoulli shift is also  present  in the orbit structure of  general covariance, even after restricting to the vacuum sector.  Our incompleteness  theorems are a  consequence of this complex orbit structure.

\medskip
\emph{Incompleteness of Observables in General Relativity---} 
We are now ready to see how Theorems \ref{T:1} and \ref{T:3} come about. The proofs follow from standard  
arguments
used in invariant descriptive set theory \cite{Gao,hjorth}. We sketch these arguments here for completeness. 

Let $\mathcal{S}$ be a rich family and fix $r\colon X\to \mathcal{S}$ as in \eqref{EQ:reduction}. Let now $f\colon \mathcal{S}\to R$ be any complete concrete observable and precompose $\widehat{f}:=f\circ r$ to get a map $\widehat{f}\colon X\to R$. 
 By \eqref{EQ:Complete} and  \eqref{EQ:reduction}, for every $\alpha,\beta \in X$ we have that:
\begin{equation}\label{EQ:FinalInvariance}
\alpha\simeq_{G}\beta \iff \widehat{f}(\alpha)=\widehat{f}(\alpha).
\end{equation}

We will make use of the  following classical result. For the proof of its first part,  see \cite[Theorem 3.2]{hjorth}  or \cite{Gao}. The second part of the statement follows directly from the first, as the  comeager $C\subseteq X$ cannot be covered by a union of finitely many meager orbits, see e.g. \cite{Kechris}. 

\begin{lemma}\label{Lemma:1}
Let  $\widehat{f}\colon X\to R$ be a Baire-measurable map which satisfies (\ref{EQ:FinalInvariance}). Then, there exists a comeager set $C\subseteq X$ on which  $\widehat{f}$ is constant.  
 In particular, there are $\alpha,\beta\in X$ with $\alpha\not\simeq_{G}\beta$ and $\widehat{f}(\alpha)=\widehat{f}(\beta)$.
\end{lemma}

 Theorem \ref{T:1}  follows  from Lemma \ref{Lemma:1}. Indeed, assume that the complete observable $f\colon \mathcal{S}\to R$ is Borel definable. It follows that the associated $\widehat{f}$ above is Baire-measurable and hence, by Lemma \ref{Lemma:1}, we have  $\alpha,\beta\in X$ with $\alpha\not\simeq_{G}\beta$ and $\widehat{f}(\alpha)=\widehat{f}(\beta)$. But this contradicts  \eqref{EQ:FinalInvariance}.

Theorem \ref{T:3} follows from Lemma \ref{Lemma:1} and the fact that  there exists, provably from  ZFC,  a model of ZF+DC in which  every map $\widehat{f}\colon X\to R$ is Baire-measurable \cite{Solovay,Shelah}. Since Lemma \ref{Lemma:1} is provable in $\mathrm{ZF}+\mathrm{DC}$, all maps in this model satisfy  the last statement of Lemma \ref{Lemma:1}, and hence, they have to fail \eqref{EQ:FinalInvariance}. Notice the resemblance  of this  proof  with the usual consistency proof   of the first four  axioms of Euclid with  the negation of the parallel postulate, which uses Euclidean geometry to construct a model of non-Euclidean geometry such as the Poincare disc.

\medskip
\emph{Discusssion---} Similar to G\"odel's first incompleteness theorem which shows that no `rich enough' fragment of arithmetic admits a consistent  extension that is both complete and computable, Theorems \ref{T:1} and \ref{T:3} show that no  `rich enough'  collection of spacetimes admits an observable that is both 
complete and definable. 

Given the central role that various  types of observables play in quantization procedures \cite{Old3,Thiemann,Car2001,ashtekar_reuter_rovelli_2015}, a natural followup question is: how much of Bergmann's  program for ``\emph{the identification and systematic exploitation of the observables}'' \cite{Quote} can be salvaged, and in what precise form? 

Some first attempts to preserve definability while trying to maintain completeness of observables on large collections of spacetimes can be  ruled out merely on the basis of how flexible is the notion of `concrete observable'  in Theorems \ref{T:1} and \ref{T:3}. This includes attempts involving gauge--fixing procedures $s\colon\mathcal{S}\to\mathcal{S} \subseteq \mathrm{Ein}(M)$
which select a single representative $s(g_{\mu\nu})\in [g_{\mu\nu}]_{\mathrm{diff}}$ from each 
class  (\ref{EQ:Abstract}); or the use of families of observables $\mathcal{F}=\{f_i\colon i\in I\}$ in place of a single observable $f$. Indeed, since $\mathrm{Ein}(M)$ is a Polish space, Theorems \ref{T:1} and \ref{T:3} imply that no gauge--fixing map $s$ can be definable. Similarly, since the notions of concreteness and definability are closed under countable products, a countable $\mathcal{F}$ can  be replaced by the single observable $\otimes_i f_i$. In fact, a technical elaboration    shows  that our incompleteness results extend  to uncountable families of observables, so long as  the parametrisation $i\mapsto f_i$ is  `definable enough'. 

One could also try  to circumvent these issues by endowing $\mathcal{S}$ with a different topology. This would have to be a topology so `fine' that it admits a Borel definable complete observable $f\colon\mathcal{S}\to \mathbb{R}$.  While this is certainly doable  --- for example one may consider the discrete topology, Theorem \ref{T:3} raises the question of whether Borel maps in this new topology would be  amenable to computations.

On the other hand it is not at all clear --- and in fact almost certainly not  a good idea ---   that  a successful theory of gravity should predicate on the definability of complete observables over large collections of spacetimes.  In fact, similar incompleteness phenomena may occur in electromagnetism,  for example, if one takes  observables to be all the Poincare-invariant quantities defined on the space of all Maxwell solutions without imposing any boundary conditions. Yet this does not seem it would cause any issues to the theory\footnote{We thank an anonymous referee for this observation.}.

We are left with many questions. How much do we need to restrict the collection of spacetimes $\mathcal{S}$ before it  admits definable complete observables?  Are there generalizations of the notion of an `observable' that allow to classify `rich' collections of spacetimes definably? Is there a formal sense in which general covariance is strictly more complex than the gauge induced, say, by actions of the Poincare group?   With such questions in mind, we next discuss two  natural directions in which this work could be extended.

\medskip

Several interesting and physically relevant collections of spacetimes   might admit definable complete observables. Consider for example the collections $\mathrm{AF}$,  $\mathrm{DCD}$, $\mathrm{VS}_{+}$,  $\mathrm{VS}_{-}$  of  all spacetimes on $\mathbb{R}^4$ which are  asymptotically flat,   maximal globally hyperbolic developments of a Cauchy data set,   and $\Lambda$-vacuum solutions with positive or negative cosmological constant $\Lambda$, respectively. 
In the context of Theorems \ref{T:1}, \ref{T:3} above it is natural to ask:

\begin{problem}\label{Problem:new}
Which of the  families above are rich?
\end{problem}

Preliminary results not provided here suggest  that $\mathrm{DCD}$ is indeed rich, but this requires some different techniques than the ones presented in this letter.  Moreover, in view of Penrose's method for approximating regions of any spacetime near a null geodesic via plane waves \cite{Penrose2} the proof for Theorem \ref{T:Examples} is likely generalizable to other families of spacetimes.

\medskip

The fully--invariant observables considered here   can be relaxed to `equivariant' types of observables. 
Let $G$ be a Polish group. A \emph{G--observable for $\mathcal{S}$} is a Polish $G$-space $G\curvearrowright R$ together with a  map $f\colon \mathcal{S}\to R$ so that 
\begin{equation}
    g_{\mu\nu}\simeq_{\mathrm{diff}} \widetilde{g}_{\rho\sigma} \iff   f(g_{\mu\nu})\simeq_G f(\widetilde{g}_{\rho\sigma})
\end{equation}
holds for all $g_{\mu\nu},\widetilde{g}_{\rho\sigma}\in\mathcal{S}$.
Instances of $G$--observables have been considered in the literature before. For example, some  modern criticisms to Bergmann's program   \cite{Pitt}  maintain the use of scalars $R=C^{\infty}(\mathbb{R}^4,\mathbb{R})$ as  values for observables, but  replace the equality in  the right--hand side of (\ref{EQ:Invariant}) with covariance $\simeq_{\mathrm{diff}}$ of scalars.

Families of spacetimes which are incomplete in the sense of Theorems \ref{T:1} and \ref{T:3}, may still admit definable $G$--observables,  for various groups $G$. If $G$ can be chosen to have nice representation-theoretic properties, then  $G$--observables can  still be promoted  to operators on a Hilbert space, and hence be used for quantization. An elaboration on Theorem \ref{T:1} shows that  rich families  $\mathcal{S}$ do not admit $G$--observables for compact $G$  \cite[Exercise 5.4.5]{Gao}. But, could $G$ be  locally--compact? Or could $G$ be the unitary group of a separable $C^*$-algebra?

\begin{problem}\label{Problem}
For which Polish groups  $G$ and  $\mathcal{S}\subseteq \mathrm{Ein}(\mathbb{R}^4)$ there exists a   definable and complete $G$--observable for $\mathcal{S}$?  
\end{problem}

Investigating Problem \ref{Problem}  may also allow to formally address whether the problem of observables is `more severe' in general relativity than in say electromagnetism,  e.g. by taking $G$ to be  the gauge  group underlying the  latter.


 Recent breakthroughs  \cite{hjorth2,Pan0,Pan1,Pan2} in the complexity theory of Polish groups actions   provide sharp tools for investigating the problems listed above. 
 We hope that this work may form the basis of a research  program for cross-pollinating theoretical physics and  descriptive set theory in order to analyze  the  complexity of general covariance.

\medskip

In summary, the roots of the problem of observables run deep and suggest another beautiful connection of mathematics with physics inspired by general relativity. This work opens many unexplored future directions. Perhaps 
further investigation  may also help identify types of quantization recipes that can be implemented definably.

\subsubsection*{Acknowledgments}We are grateful to Jonathan Holland for sharing with us his intuitions on the symmetries of plane waves. We are also grateful to the anonymous referees who helped us improve the scope of the paper significantly. 
We thank Apoorv Tiwari for many valuable discussions in the early stages of this project, Hans Halvorson for the phrase `when an object exists only by the power of AC then for what concerns physics it is as useful as if it did not exist', Carlo Rovelli and Jeremy Butterfield for commenting on a draft of the manuscript. Finally, we would like to thank Clemmie Murdock III for introducing AP to GS.

This research was supported by the
NSF Grant DMS-2154258: ``Dynamics Beyond Turbulence and Obstructions to Classification''. MC acknowledges support from the ID\# 61466 and ID\# 62312 grants from the John Templeton Foundation, as part of the ``Quantum Information Structure of Spacetime (QISS)'' project (\hyperlink{http://www.qiss.fr}{qiss.fr}).

\bibliography{references}

\clearpage

\thispagestyle{empty} 
\section{Supplementary Material}

Recall the relation $\simeq_{\mathbb{Z}}$ on  $X:=\{0,1\}^{\mathbb{Z}}$ given by \eqref{EQ:ZShift}. Here we start by  providing the exact definition of the  map $r\colon X\to \mathrm{GPW}$ and the details as to why it is a Borel reduction from $(X,\simeq_{\mathbb{Z}})$ to $(\mathrm{GPW}, \simeq_{\mathrm{diff}})$. We then sketch how this argument adapts for showing that the family of  all Robertson-Walker spacetimes is also rich.

\begin{figure}[h!]
\includegraphics[scale=0.25]{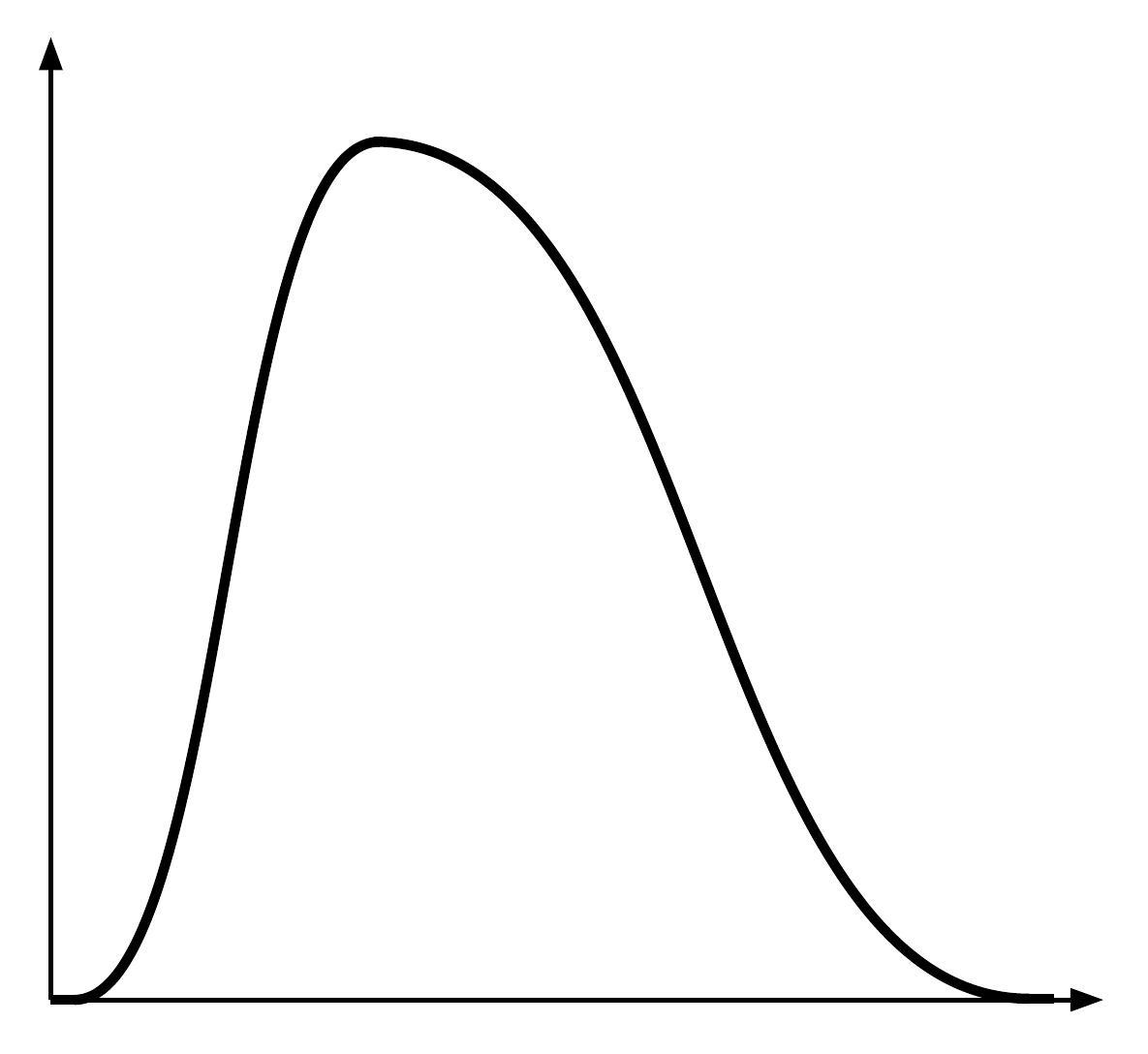} 

\caption{The ``bell curve" $W$}
\label{Fig:2}
\end{figure}

\medskip
\emph{Gravitational Plane Waves---}
Let $W\colon[0,1]\to [0,1]$ be any smooth map so that: $W$ and all its derivatives vanish at $x=0$ and $x=1$; and $W(x)= 1$ for some unique point in $[0,1]$   with $x<1/2$. 
For  every element $\alpha\colon\mathbb{Z}\to \{0,1\}$ of $X$ we define a smooth map $W_{\alpha}\colon\mathbb{R}\to\mathbb{R}$ by setting: 
\begin{equation}
W_{\alpha}(x)=\begin{cases*}
                    2-W(x-\lfloor x \rfloor) & \text{if}  $\alpha(\lfloor x \rfloor ) =0$ \\
                    2+W(x-\lfloor x \rfloor)  & \text{if} $\alpha(\lfloor x \rfloor ) =1$ 
                 \end{cases*} 
\end{equation}
For the graphs of $W_{\alpha}$, $W$ see Figures \ref{Fig:1}, \ref{Fig:2}, respectively.

Recall that for every $\alpha\in X$, the spacetime $r(\alpha)$ is defined by setting   $H(u,x,y):=W_{\alpha}(u)xy$ in  (\ref{EQ:PlaneWaves}). To see that  $r$ is continuous, and hence Borel,  notice that compact regions of $r(\alpha)$ are determined by finite regions of $\alpha$, and that basic open sets in $X$ are of the form:
\begin{equation}\label{EQ:Open}
U_{p}:=\{\alpha \in X \text{ with } \alpha \upharpoonright \mathrm{dom}(p) =p \},
\end{equation}
where $p\colon A\to\{0,1\}$ is a map with finite  $A\subseteq \mathbb{Z}$.  It is also straighforward to see that  $r$ satisfies the $(\Longrightarrow)$ direction of (\ref{EQ:reduction}). Indeed, if $\alpha\simeq_{\mathbb{Z} }\beta$ then there is some  $k\in\mathbb{Z}$  so that for all $u\in \mathbb{R}$ we have that $W_{\alpha}(u-k)=W_{\beta}(u)$. But then, the chance of coordinates $u\mapsto u-k$, $v,x,y\mapsto v,x,y$ witness that $r_{\alpha}\simeq_{\mathrm{diff}} r(\beta)$. To prove that the    ($\Longleftarrow$) of (\ref{EQ:reduction})   holds as well, we will need to recall some basic facts regarding Killing symmetries, see for instance \cite[Chapter 8]{SKMHH}.

For any spacetime $g_{\mu\nu}$, let $\mathfrak{iso}_{\mathcal{L}}(g_{\mu\nu})$ be the Lie algebra of all the vector fields $\boldsymbol{V}$ which satisfy Killing's equation:
\begin{equation}\label{EQ:Killing}
\mathcal{L}_{\boldsymbol{V}} g_{\mu\nu}:= g_{\mu\nu,\alpha} \; V^{\alpha}+g_{\alpha\mu} \; V^{\alpha}_{\; ,\nu} + g_{\alpha\nu} \; V^{\alpha}_{\; ,\mu}=0
\end{equation}
We will make repeated use of the  standard fact that, if
$\widetilde{x}^{\xi}=\widetilde{x}^{\xi}(x^{\eta})$ is a smooth change of coordinates  witnessing  $g_{\mu\nu}\simeq_{\mathrm{diff}}\widetilde{g}_{\rho\sigma}$ via (\ref{EQ:GC}), then it also induces an isomorphism  $i\colon \mathfrak{iso}_{\mathcal{L}}(g_{\mu\nu}) \to  \mathfrak{iso}_{\mathcal{L}}(\widetilde{g}_{\rho\sigma})$, with $i(\boldsymbol{V})=\boldsymbol{\widetilde{V}}$, where
\begin{equation}\label{EQ:VectorFields}
V^{\alpha}(x^{\eta})=  \frac{\partial  x^{\alpha} }{\partial \widetilde{x}^{\beta}} \widetilde{V}^{\beta}(\widetilde{x}^{\xi}). 
\end{equation}

We can now prove that the  ($\Longleftarrow$) direction of (\ref{EQ:reduction}) holds:

\begin{lemma} \label{L:2}
If  $r(\alpha)\simeq_{\mathrm{diff}} r(\beta)$ holds then so does $\alpha \simeq_{\mathbb{Z}} \beta$.
\end{lemma}
To see this, one  starts by solving (\ref{EQ:Killing})  for an arbitrary vector field $\boldsymbol{V}$ and for any $g_{\mu\nu}$ of the form (\ref{EQ:PlaneWaves}). Solutions for this can be found in 
\cite[Chapter 4.3]{JEK}, for the case $H_{xx}-H_{yy}=0$ in which we are interested; and in \cite{Sippel}, for the general $H(u,x,y)$. 
 The additional requirement that  $H(u,x,y)=W_{\alpha}(u)xy$, for some $\alpha\in X$, implies that  $\mathfrak{iso}_{\mathcal{L}}(g_{\mu\nu})$ is the $5$-dimensional Heisenberg algebra $\mathfrak{h}(2)$; see \cite[4.3.16]{JEK} or  \cite[Table II.10]{Sippel}.
In particular, the  center of  $\mathfrak{iso}_{\mathcal{L}}(g_{\mu\nu})$  is spanned by $\frac{\partial}{\partial v}$.
    As a consequence, if (\ref{EQ:GC}) holds for some $\widetilde{x}^{\xi}=\widetilde{x}^{\xi}(x^{\eta})$ and $g_{\mu\nu}:=\rho(\alpha), \widetilde{g}_{\rho\sigma}:=\rho(\beta)$, then it should map  $\partial/\partial \widetilde{v}$ to a constant multiple of  $\partial/\partial v$. In particular, $\partial u/\partial \widetilde{v}=0$, and hence, $\widetilde{x}^{\xi}=\widetilde{x}^{\xi}(x^{\eta})$ is of the following form; see, \cite[(A.1)]{Sippel} or \cite[4.3.1]{JEK}:  
    
\begin{align}
\begin{split}\label{EQ:Final}
  \widetilde{u} &=  \frac{u+a}{c} \\
 \widetilde{x} &= x \cos(b) + y \sin (b) + F(u) \\
  \widetilde{y}& = -x \sin(b) + y \cos (b) + G(u) \\
  \widetilde{v} &= c \big[ v- x\big(\cos(b)F'(u)-\sin(b) G'(u)\big) \\ 
                      &\quad  -y\big(\sin(b)F'(u)-\cos(b)G'(u)\big)-I(u)\big]. 
\end{split}
\end{align}
for some constants $a,b,c$, with $c\neq 0$, and some smooth maps $F(u),G(u),I(u)$. Plugging (\ref{EQ:PlaneWaves}) and (\ref{EQ:Final}) in  (\ref{EQ:GC}) and solving for the coefficients of $xydu^2$ we get:
\begin{equation*}
W_{\alpha}(u)=\frac{\cos^2(b)-\sin^2(b)}{c^2}\cdot W_{\beta}(\frac{u+a}{c})
\end{equation*}
 From the structure of mimima and maxima of $W_{\alpha}$ and $W_{\beta}$, and since
 $W$ in Figure \ref{Fig:2} is not symmetric under any vertical axis, it follows that $W_{\alpha}(u)= W_{\beta}(u+a)$ holds for some $a\in\mathbb{Z}$.  Thus, $\alpha(n)=\beta(n-k)$ for  $k:=-a$.

\medskip
\emph{Robertson-Walker cosmological spacetimes---} 
For any  $d>0$, let $\mathrm{RW}(\mathbb{R}^{d+1})$ be the collection of all $(d+1)$--dimensional Robertson-Walker spacetimes:
\begin{equation}\label{EQ:FLRW}
- dt^2+J(t)\big((dx^1)^2+\cdots +(dx^d)^2 \big),
\end{equation}
where  $J$   is a smooth map with $J(t)>0$; see, e.g., \cite[Chapter 8]{Cosmo}).  By an argument similar to the above one may show that  $\mathrm{RW}(\mathbb{R}^{d+1})$ is rich, and hence, it satisfies the conclusions of our Incompleteness Theorems \ref{T:1} and \ref{T:3}.

Indeed consider the map $r\colon X \to \mathrm{RW}(\mathbb{R}^{1+d})$ given by setting   $r(\alpha)$ to be of the form (\ref{EQ:FLRW}), with $J(t):=W_{\alpha}(t)$.
As in the case of $\mathrm{GPW}$, one easily shows that $r$ is continuous and that it satisfies  the $(\Longrightarrow)$ direction of (\ref{EQ:reduction}). The analogue of Lemma \ref{L:2} is proved in a similar fashion.
One starts by solving (\ref{EQ:Killing}); the additional requirement that  $J(t)=W_{\alpha}(t)$ implies that  $\mathfrak{iso}_{\mathcal{L}}(g_{\mu\nu})$ is the special Euclidean algebra  $\mathfrak{iso}(d)$ corresponding to the isometries of the spacelike surfaces $t=\text{constant}$.
Assuming that  (\ref{EQ:GC}) holds for  $\widetilde{x}^{\xi}=\widetilde{x}^{\xi}(x^{\eta})$,  $g_{\mu\nu}:=r(\alpha),  
\widetilde{g}_{\rho\sigma}:=r(\beta)$, one gets an isomorphism between  $\mathfrak{iso}_{\mathcal{L}}(g_{\mu\nu})$  and $\mathfrak{iso}_{\mathcal{L}}(\widetilde{g}_{\rho\sigma})$, which restricts to an isomorphism between the subalgebras  $\langle\partial/\partial x^{\eta} \colon \eta\neq 0\rangle$  and  $\langle\partial/\partial \widetilde{x}^{\xi} \colon \xi\neq 0\rangle$. It follows that there are
 smooth maps $F^{0},\ldots,F^{d}$ and constants $c_{\;\eta}^\xi$ with:
\begin{equation}\label{EQ:N2}
\widetilde{x}^0=F^0(x^0), \;  \text{ and } 
\; \widetilde{x}^{\xi}=F^{\xi}(x^0)+\sum_{\eta>0} (c_{\;\eta}^\xi \cdot x^{\eta}).
\end{equation} 
Plugging (\ref{EQ:N2}) in  (\ref{EQ:GC}) and solving for the coefficients of each combination
$dx^{\eta}dx^{\eta'}$ we get $\partial F^{0} / \partial x^0 =\pm 1$. From the structure of mimima and maxima of $W_{\alpha}$ and $W_{\beta}$, and since
 $W$ in Figure \ref{Fig:2} is not symmetric under any vertical axis, we have that  $F^0\colon \mathbb{R}\to \mathbb{R}$ is the shift map $t\mapsto t-k$ for some $k\in\mathbb{Z}$. Hence, $\alpha(n)=\beta(n-k)$ for some $k\in\mathbb{Z}$.

\end{document}